\documentclass[]{spie}  

\usepackage{amsmath,amsfonts,amssymb}
\usepackage{graphicx}
\usepackage{soul}
\usepackage[square,sort,comma,numbers]{natbib}
\usepackage[colorlinks=true, allcolors=blue]{hyperref}

\usepackage{array}

\usepackage{aas_macros}
\usepackage[bottom]{footmisc}

\usepackage[normalem]{ulem}         
\usepackage{xcolor}
\definecolor{green2}{rgb}{0.0,0.5,0}

\title{Precision speckle interferometry with CMOS detector}

\author[a]{Ivan Strakhov}
\author[a]{Boris Safonov}
\author[a]{Dmitry Cheryasov}

\affil[a]{Sternberg Astronomical Institute, Lomonosov Moscow State University, 119992 Universitetskii prospekt 13, Moscow, Russia}

\authorinfo{Further author information: (Send correspondence to I.S.)\\E-mail: strakhov.ia15@physics.msu.ru}
\begin{document}
\maketitle

\begin{abstract}
Speckle polarimeter (SPP) is a facility instrument of the 2.5-m telescope of the Caucasian Mountain Observatory of SAI MSU. By design it is a combination of a speckle interferometer and a dual--beam polarimeter. In 2022 we performed a major upgrade of the instrument. New version of the instrument features Hamamatsu ORCA-Quest qCMOS C15550-20UP, having subelectron readout noise, as a main detector, as opposed to EMCCD Andor iXon 897 used in previous version. Optical distortions present in the instrument are considered as they directly affect the accuracy of the speckle interferometric astrometric measurements of binary stars. We identified the Atmospheric Dispersion Compensator (ADC) as the main source of distortions which are not constant and depend on the rotational angles of ADCs prisms. Distortions are estimated using internal calibration light source and multiple binary stars measurements. Method for their correction is developed. Flux ratio estimates are subject to CMOS-specific negative factors: spatially correlated noise and flux--dependent pixel--to--pixel sensitivity difference. We suggest ways to mitigate these factors. The use of speckle transfer function measured using a reference star further improves flux ratio estimation performance. We discuss the precision of the estimates of position angle, separation and flux ratio of binary stars. 
\end{abstract}


\section{Introduction}
\label{sec:intro}

In the age of adaptive optics systems and Gaia, speckle interferometry remains highly relevant in the field of binary stars research. Indeed while adaptive optics provide deep contrast and operate in infrared, making it sensitive to faint stellar and planetary companions, speckle interferometry is more productive in terms of time spent on one object thanks to smaller overheads: $\gtrsim15-30$ objects can be observed per hour \citep{Tokovinin2018}. This kind of efficiency is valuable when there is need to survey moderately large number of objects for binarity. 

For example, in Kepler and TESS photometry, due to large angular size of a pixel, the flux from nearby stars can contaminate the flux from exoplanet host and consequently bias the transit depth measurements \citep{Ciardi2015}. In order to correct for this bias one should know the fluxes from the contaminating stars. When the latter are located closer than $\approx1^{\prime\prime}$ to the exoplanet host star Gaia cannot resolve them \citep{Ziegler2018}. The use of diffraction limited imaging at a large optical telescope can partially remedy this problem and expand the parameter space in which a stellar companion can be detected to separations of 20-50~mas \citep{Ziegler2020,Scott2021}.

At the moment of writing Exoplanet Follow-up Observation Program website\footnote{\href{ExoFOP}{https://exofop.ipac.caltech.edu/tess/} DOI: 10.26134/ExoFOP5} contained 30088 high angular resolution observations of exoplanet host stars. 62.7\% of these observations were conducted by means of speckle interferometry while adaptive optics imaging amounts to 29.3\%, what highlights the importance of speckle interferometry.

At 2.5-m telescope of Sternberg Astronomical Institute (SAI-2.5m) speckle interferometry is implemented using SPeckle Polarimeter (SPP), an instrument combining the features of a speckle interferometer and of a dual--beam polarimeter with a rotating half--wave plate. In period 2015--2022 we used a EMCCD Andor iXon 897 as a main detector. In July--August 2022 we upgraded the instrument, replacing EMCCD with CMOS Hamamatsu ORCA-Quest qCMOS C15550-20UP. Another important change is related to polarimetric optics. Since 2022 we use a polarization beam splitter cube instead of a Wollaston prism as an analyzer.

SPP is actively being used for observations of binary stars \citep{Knudstrup2022,Rodriguez2023,Belinski2022}, the processing produces  separation, position angle and relative flux of components. While there is an obvious motivation to detect fainter and closer components, there is another important aspect of efficiency of speckle interferometric measurements: precision of relative astrometric and photometric measurements. Multi--epoch high--precision astrometry can be used to constrain orbits of long--period visual binaries and detect multiplicity \citep{Tokovinin2016,Tokovinin2018}. Precise knowledge of flux ratio is important for study of statistics of mass ratio of binary stars \citep{El-Badry2019,Malofeeva2023}. 

Multiple epoch observations in 2015--2022 with SPP revealed that typical reproducibility of separation and position angle are 0.6\% and $0.5-1.0^{\circ}$, respectively. For relative flux the situation is worse, it demonstrates a correlation with seeing conditions: better seeing produced smaller flux ratios. Here we discuss possible factors degrading astrometric and photometric accuracy and ways to mitigate them.

\begin{figure}
\vspace{-0.1cm}
\begin{center}
\includegraphics[width=0.85\linewidth, bb= 80 450 7900 5400,clip]{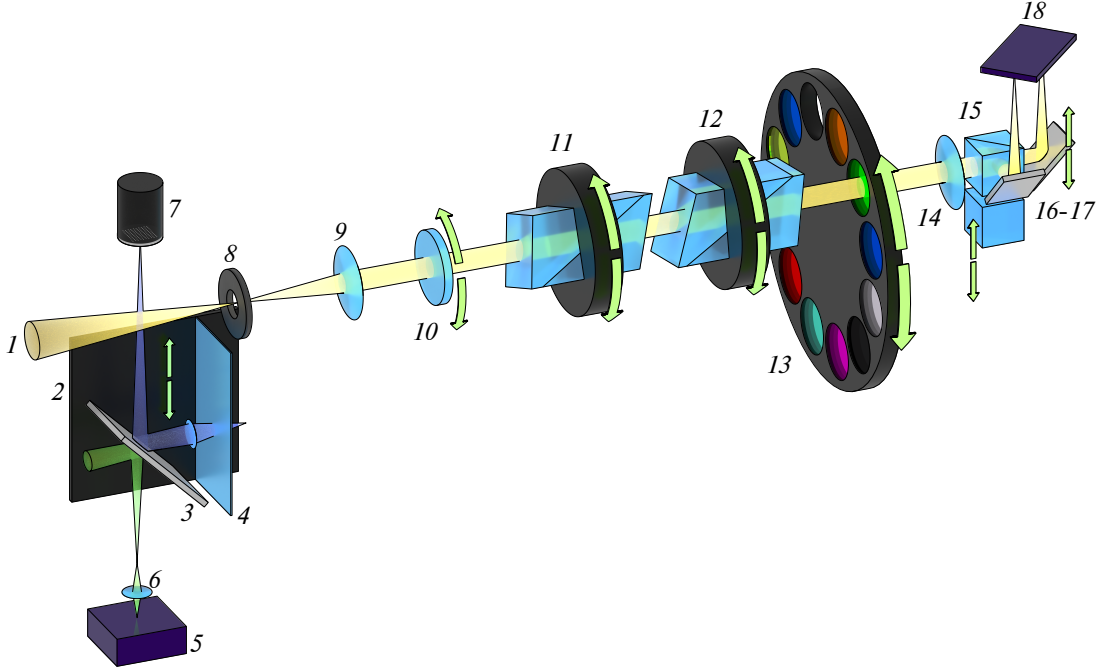}
\caption{Instrument layout. {\it 1}---Beam from the telescope. The prefocal unit {\it2--8}: the movable carriage {\it 2}, 
mirrors {\it 3}, one of the mirrors relays  light from the telescope to the auxiliary camera ({\it5--6}), another mirror relays light from the calibration source ({\it7}) to the detector, {\it4}---the linear polarizer, {\it8}---the field aperture; the collimated beam unit {\it9--14}: {\it9}---the collimator, {\it10}---the rotating half-wave plate, {\it11, 12}---the prisms of the ADC (the wedge angles are increased for clarity), {\it13}---the filter wheel, {\it14}---the lens; the camera unit {\it15--18}: {\it15}---the beamsplitter, {\it16, 17}---the relay mirrors, {\it18}---the main detector. The green arrows mark the motorized degrees of freedom. }
\label{fig:scheme}
\end{center}
\end{figure}

\section{CMOS vs EMCCD}
\label{sec:cmos}
EMCCDs have several features making them ideal detectors for methods operating with large number of short--exposure images. They have sub--electron readout noise, high readout speed, high quantum efficiency. The drawbacks should be mentioned as well. The number of pixels is modest. Subelectron level of readout noise is reached at high values of EM gain, which in turn leads to so called multiplication noise, effectively doubling photon noise. Another negative factor is very low dynamic range, typically one pixel can contain only hundreds of photoelectrons at high EM gain. Clock induced charge creates background which is critical when observing faint targets.

Recently CMOS technology matured enough to challenge the dominance of EMCCDs. The readout noise of commercially available detectors, e.g. Hamamatsu ORCA-quest C15550-20UP is lower than 1 electron without application of EM Gain, therefore without drawbacks associated with it. Therefore we decided to use this detector as a main detector in SPeckle Polarimeter. Recently the same detector was installed in upgraded SCExAO/VAMPIRES instrument at Subaru\footnote{Paper 13096-110 ``The next generation of high-contrast visible polarimetry with SCExAO/VAMPIRES'' by Lucas, M., this conference.}. We expect expansion of use of such detectors in astronomy, especially for the implementation of methods requiring obtaining series of short--exposure images.

Simple analytical considerations, detailed simulations and practical measurements with SPP in 2022--2024 have shown that particular CMOS indeed performs better than particular EMCCD. Gain in terms of SNR in power spectrum for faint objects is 2--3 \citep{Strakhov2023}. However we had to handle correlated noise which was absent in EMCCD. The procedure is similar to one developed by \citep{Schlawin2020}. The horizontal and vertical structure quasi--static structure in readout noise was estimated in areas not containing the scientific object and subtracted.

\section{Astrometry}
\label{sec:astro}

\begin{table}[b]
\caption{Binary stars used for calibration procedure. $N_{obs}$ total -- total number of observations, $N_{obs}$  ``auto'' -- number of observations conducted with ADC ``auto'' mode. $N_{obs}$  ``off'' -- number of observations conducted with ADC ``off'' mode. $\theta$, $\rho$ -- position angle and separation of binary. A Gmag, B Gmag -- Gaia magnutide of binary components. $\rho$ and $\theta$ values are given for 2023.5 epoch for illustrative purposes. The calibration procedure takes the epoch of each observation into account.
\label{table:calib_stars}}
\begin{center}
\setlength\extrarowheight{1pt}
\begin{tabular}{|l|c|c|c|c|c|c|c|}
\hline
                 & \multicolumn{3}{c|}{$N_{obs}$} &  &  & \multicolumn{2}{c|}{Gmag}  \\
\cline{2-4}
\cline{7-8}
WDS identifier   & total & ``auto'' & ``off'' & $\rho, {}^{\prime\prime}$ & $\theta, {}^{\circ}$ & A & B \\\hline
WDSJ14497+4843AB & 13         & 6                  & 7                 & 2.56                      & 45.90                & 6.12 & 6.52 \\
WDSJ23595+3343AB & 2          & 1                  & 1                 & 2.61                      & 343.95               & 6.34 & 6.54 \\
WDSJ04335+1801AB & 3          & 2                  & 1                 & 3.06                      & 276.83               & 6.89 & 7.00 \\
WDSJ00548+0926AB & 2          & 1                  & 1                 & 3.11                      & 299.54               & 8.48  & 9.45 \\
WDSJ01554+7613AB & 4          & 2                  & 2                 & 3.16                      & 242.26               & 7.34 & 8.12 \\
WDSJ00527+6852AB & 2          & 1                  & 1                 & 3.18                      & 39.99                & 7.91 & 7.97  \\
WDSJ17512+4454AB & 2          & 1                  & 1                 & 3.38                      & 326.55               & 7.99 & 8.01 \\
WDSJ08508+3504AB & 14         & 8                  & 6                 & 3.45                      & 278.66               & 7.43 & 7.50 \\
WDSJ03171+4029AB & 14         & 8                  & 6                 & 3.62                      & 29.80                & 6.80 & 7.81 \\
WDSJ12043+2128AB & 5          & 3                  & 2                 & 3.69                      & 235.28               & 6.11 & 7.49 \\
WDSJ05499+3147AB & 10         & 6                  & 4                 & 3.70                      & 61.80                & 7.20 & 8.18 \\
WDSJ02581+6912AB & 3          & 2                  & 1                 & 4.10                      & 83.09                & 7.78 & 9.82 \\
WDSJ06482+5542AB & 9          & 4                  & 5                 & 4.47                      & 76.23                & 6.16 & 6.20 \\
WDSJ11024+8313AB & 2          & 1                  & 1                 & 4.51                      & 25.11                & 8.80 & 9.70 \\
WDSJ02292+5352AB & 19         & 9                  & 10                & 4.68                      & 35.53                & 8.70 & 9.53 \\
WDSJ04320+5355AB & 11         & 7                  & 4                 & 10.36                     & 308.14               & 5.71 & 6.92 \\

\hline
\end{tabular}
\end{center}
\end{table}

Speckle interferometric astrometric measurables are position angle ($\theta$) and separation ($\rho$). Among potential sources of errors are unaccounted instrumental distortions and aberrations, imperfections of telescope mount and optics alignment (e.g. collimation error which is the non-zero angle between the optical axis and the plane perpendicular to the elevation axis), unaccounted errors in reference catalogue.

The calibration procedure of position angle and separation is following. We observe multiple known wide ($\rho\sim2\div10^{\prime\prime}$) binaries  (see Table~\ref{table:calib_stars}). The amount of 3000 to 12000 short-exposure ($\mathrm{exp}\approx22$~ms) frames is accumulated during one observation of the binary. At the first step of the processing we get the initial rough estimate from the positions of maxima on shift-and-add averaged image. Based on these rough estimates we crop out $160\!\times\!160$ px window around both components on every frame of the series and use one of the windows as the reference point spread function $T_i$. Using the window around the other component $I_i$ we now apply the speckle deconvolution method \citep{Primot1990}:
\begin{equation}
\widetilde{O}_e = \frac{\langle \widetilde{I_i}\widetilde{T_i}^* \rangle}{\langle |\widetilde{T_i}|^2 \rangle},
\end{equation}
where $\widetilde{.}\,$ symbol denotes Fourier transform, $i$ is the frame index, operator $\langle . \rangle$ means the average over all frames of the series, $\widetilde{O}_e$ is the estimated object visibility. We approximate the phase of the visibility function $\mathrm{Arg}\left(\widetilde{O}_e\right)$ by the plane (under the assumption that both components are point--like sources). The subpixel refinement values for the initial estimates of separation and position angle are obtained from the plane slope value and the direction of the slope. 

Next, we compare the obtained $\rho$ and $\theta$ to the predicted values from Gaia DR3. For this procedure we select only the observations with relatively small estimated position angle errors ($<\!0.045^{\circ}$). The number of selected observations is 115, the number of unique binaries is 16. The comparison yields the correction value for the position angle and the angular scale both averaged over all observed binaries estimations.  Our objective is to reduce the dispersion of the data over all observed binaries. We need to consider the existence of the systematic error sources during this procedure. Thus, at this step we can take into account the collimation error term, it can be estimated from the same data and equals $190^{\prime\prime}$. The neglect of this factor leads up to $0.8^{\circ}$ unaccounted position angle offset if observation is conducted at high altitudes. This term only affects the position angle estimation.

\begin{figure}[t!]
\begin{center}
\includegraphics[width=0.8\linewidth]{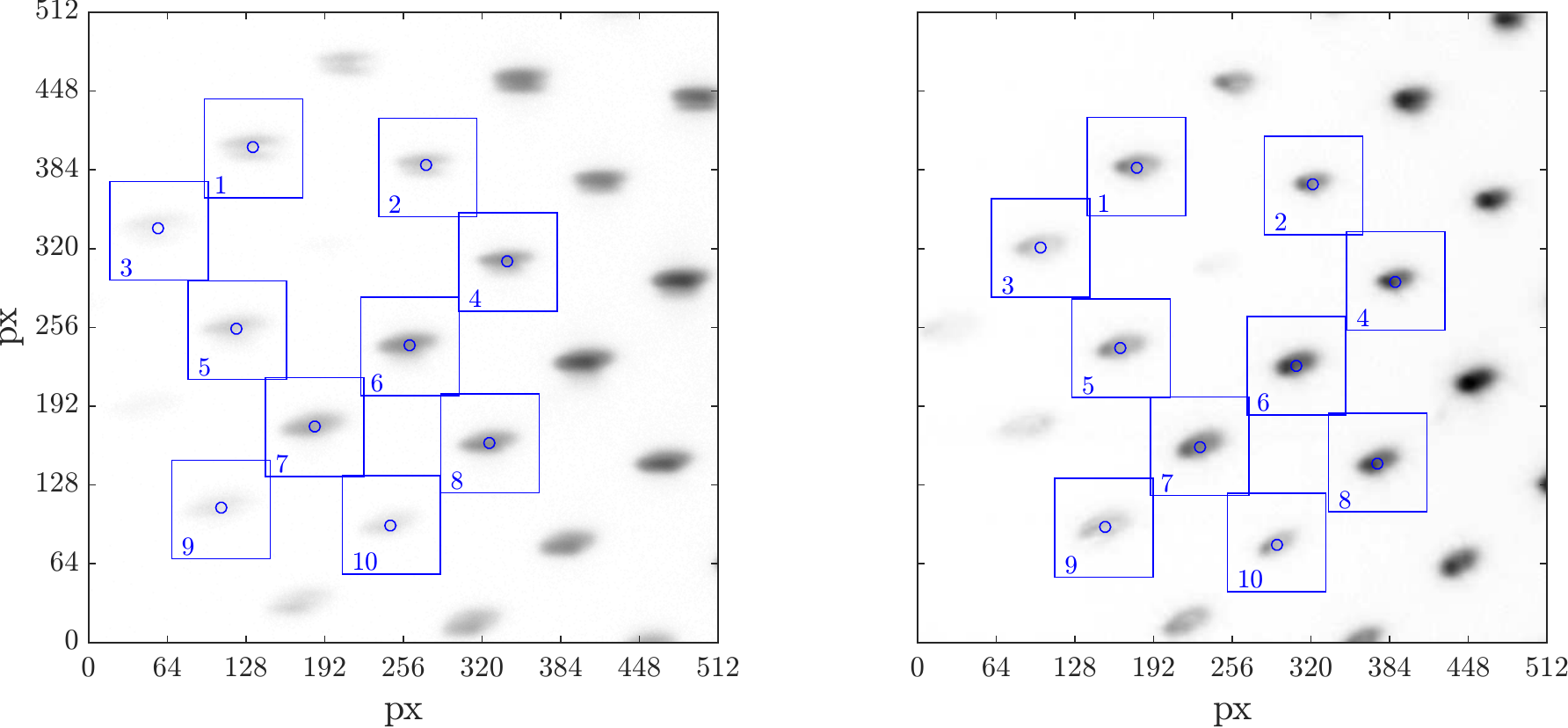}
\caption{Example of calibration source images used in distortion correction procedure. Left image obtained with ADC, right image obtained without ADC. Blue squares denote the positions of source points used to determine affine geometric transformation from distorted to undistorted image. The position of the spot is estimated as its center of mass. Images were obtained through the filter centered on 550~nm and 50~nm wide. The spots appear elongated due to dispersion introduced by ADC.}
\label{fig:adc_calib_source}
\end{center}
\end{figure}

\begin{figure}[t!]
\begin{center}
\includegraphics[width=0.8\linewidth]{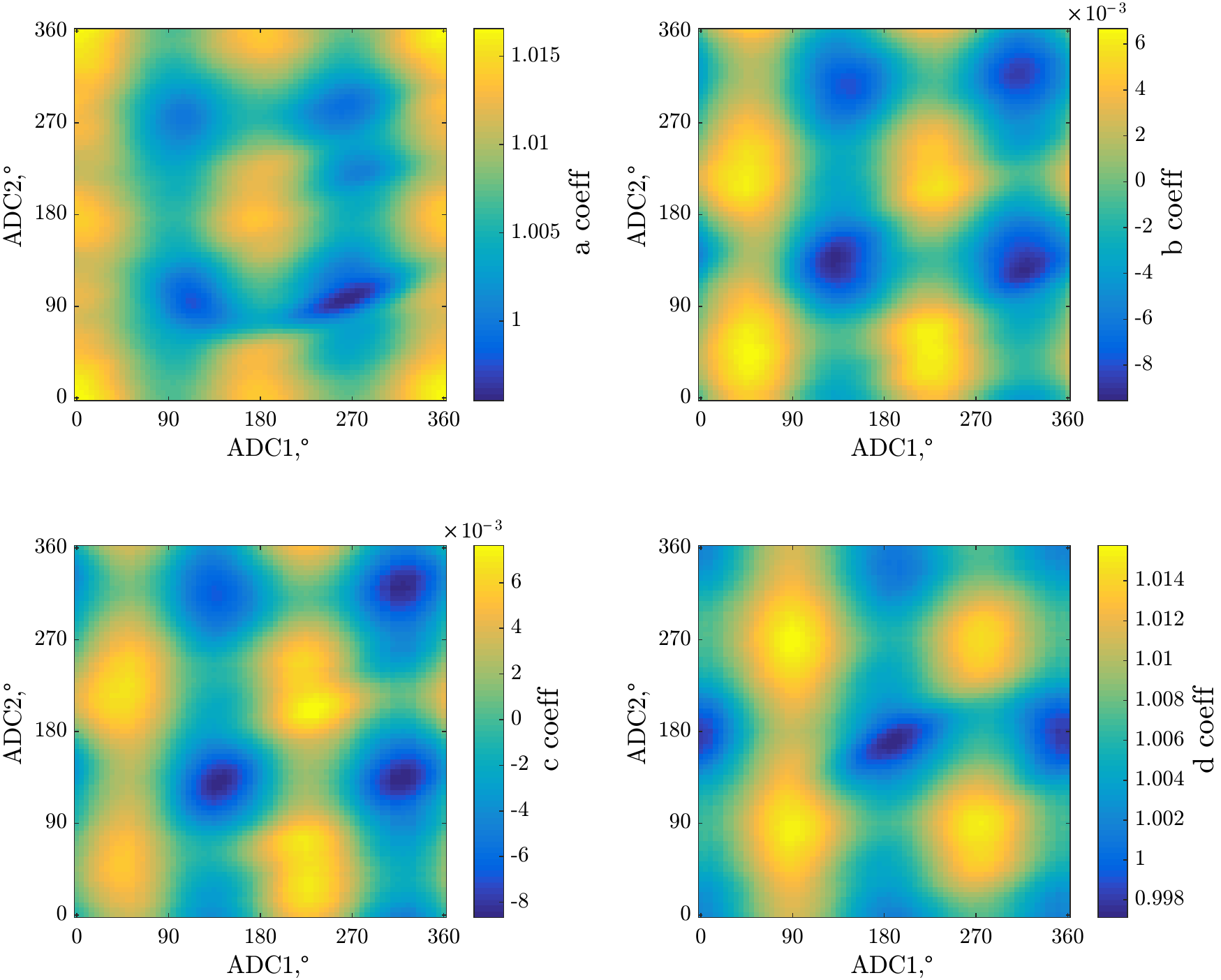}
\caption{Affine transform coefficients for every position of ADC prisms.}
\label{fig:adc_interf}
\end{center}
\end{figure}

At the same time, the angular scale estimate is also subject to systematic effects. We identify the distortions introduced by atmospheric dispersion compensator (ADC) prisms as the main source of these effects. ADC is installed in the collimated beam after the half-wave plate (see Fig.~\ref{fig:scheme}). ADC is represented by the two direct vision prisms installed in two independently rotation stages. Each prism, in turn, consists of two prisms made of LZOS F1 and LZOS K8 glass, with wedge angles $15\,.\!\!^\circ9$ and $18\,.\!\!^\circ65$, respectively. The aperture of the prisms is \mbox{$15\times15$}~mm. 

We measured distortions introduced by ADC using an internal light source which is presented by multiple holes on a grid $10\!\times\!10$ projected onto an image plane. The image positions (centers of mass) of each dot were registered with and without ADC in the light beam (see Fig.~\ref{fig:adc_calib_source}). When ADC was introduced, dots images positions were registered for each of the prisms angles combination (with 5 degree step for each prism and 10 degree step for angle between prisms). In total 2592 combinations were considered.

To describe the distortion we consider the 2D affine transformation in the following form:
\begin{equation}
    \label{eq:OTF}
    \begin{bmatrix}
x & y & 1\\
\end{bmatrix}=\begin{bmatrix}
u & v & 1\\
\end{bmatrix}\begin{bmatrix}
a & b & 0\\
c & d & 0\\
t_x & t_y & 1\\
\end{bmatrix},
\end{equation} where we transform the point $(u,v)$ in the original coordinate space to the point $(x,y)$ in the output coordinate space, $a,b,c,d$ are scale and shear coefficients we are interested in, $t_x, t_y$ are the translation coefficients not affecting angle or scale and, therefore, are of no interest to us. The affine geometric transform was estimated for each ADC state to reduce it to the configuration without ADC. As a result, we obtained the affine tranformation $a,b,c,d$ coefficients for each ADC state (see Fig.~\ref{fig:adc_interf}) by linearly interpolating the in-between points. From now on we can use this matrix in calibration and speckle interferometric data reduction to mitigate distortion effects.  

To verify this approach we observed every binary from Table~\ref{table:calib_stars} in two states of ADC: state ``off'' and state ``auto'' (observations were conducted immediately one after another). In state ``off'' position angles of the first and second ADC prism were set to $0^{\circ}$ and $180^{\circ}$, respectively. In this state the ADC does not introduce any dispersion and therefore atmospheric dispersion stays uncompensated. State ``auto'' means that angles of ADC prisms are set so that the ADC produces dispersion equal to atmospheric one but in opposite direction, effectively compensating the latter. The states differ in ADC prisms positions, hence the introduced distortions are different as well.
First, we execute the calibration procedure without ADC distortion correction and compare angle and scale estimates between the two states. The result of the comparison is following: position angles estimates differ by $0.2-0.3^{\circ}$ and scale estimates differ by $0.05-0.1$ mas/px. At the same time, by the means of ADC distortion correction we reduce the differences between the two states to $0.05^{\circ}$ and $0.02-0.05$ mas/px respectively.

\begin{figure}[h]
\begin{center}
\includegraphics[width=1.0\linewidth]{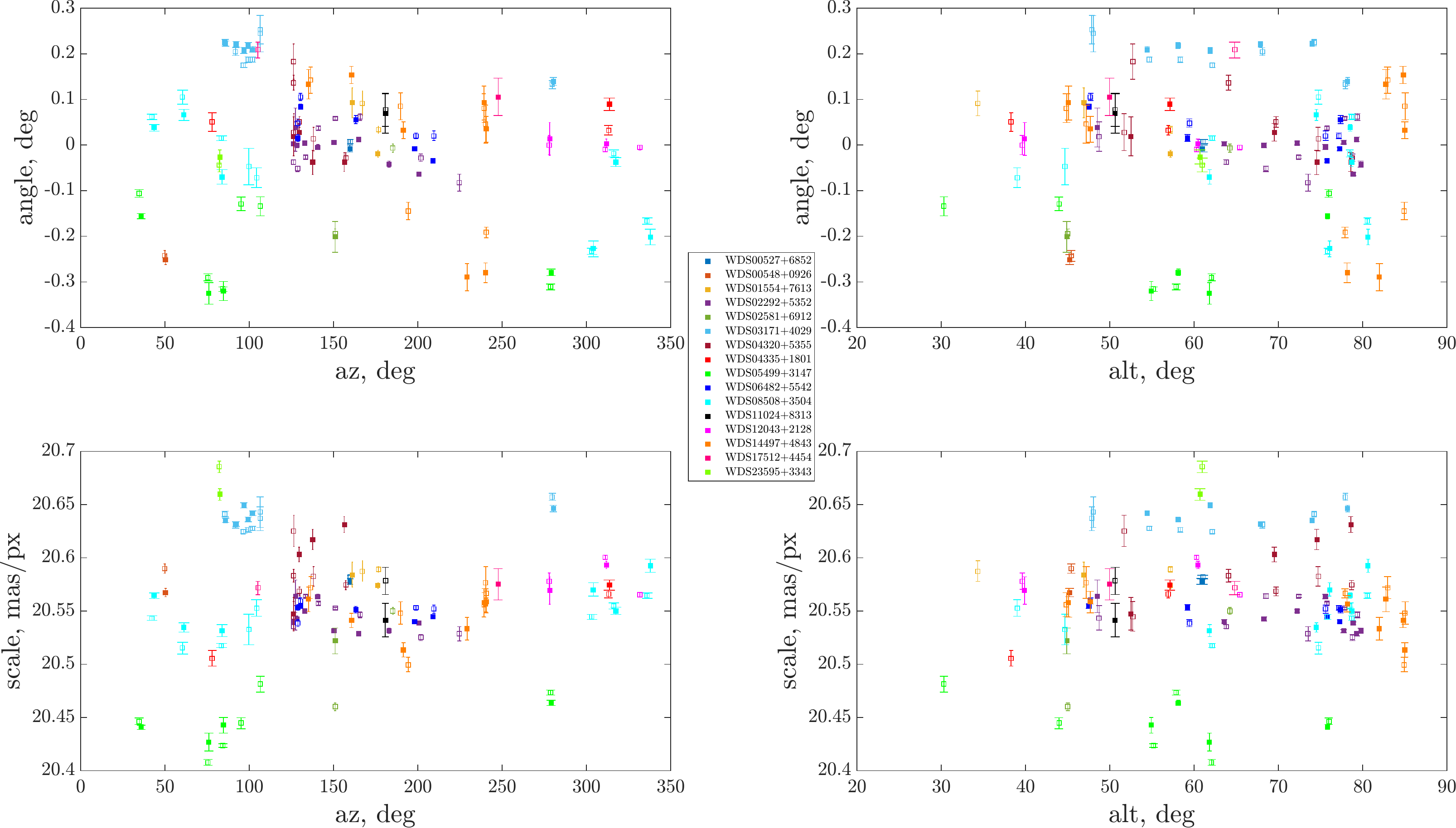}
\caption{Calibration binary observations after the reduction of collimation error and ADC distortion. Filled squares -- ``off'' ADC state, empty squares -- ``auto'' ADC state.}
\label{fig:astrometry}
\end{center}
\end{figure}

\begin{figure}[h]
\hspace{0.19cm}
\includegraphics[width=0.9785\linewidth]{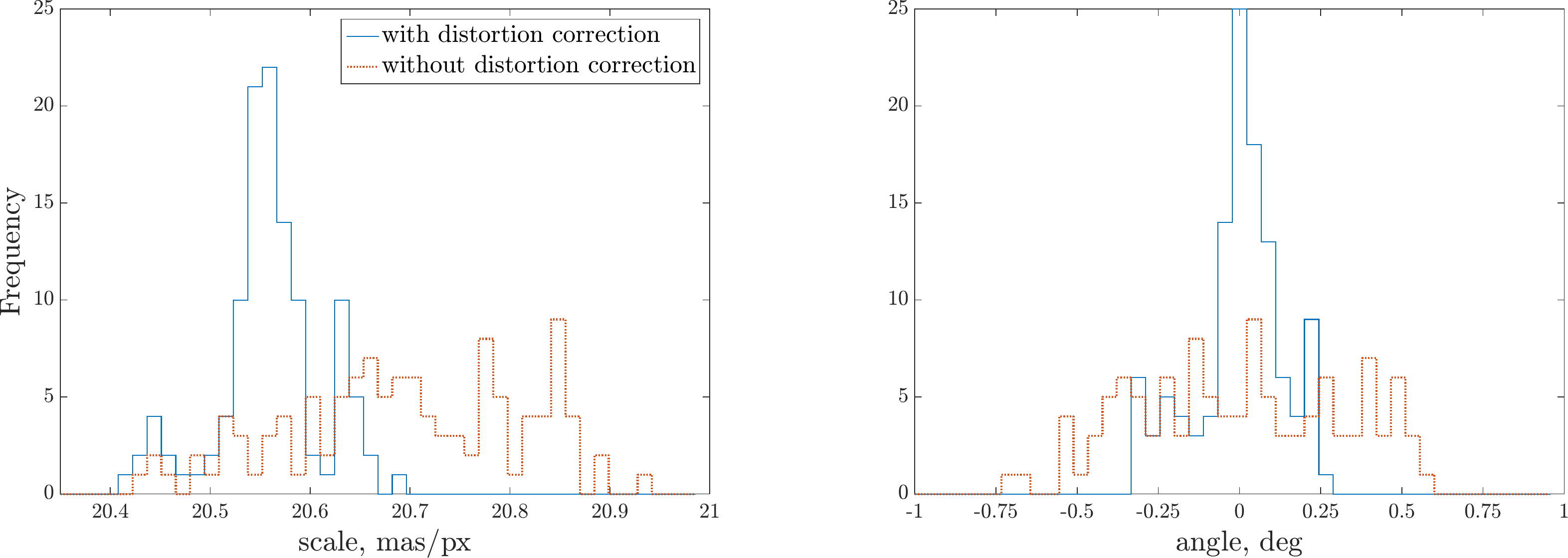}
\caption{Histograms of calibration binary measurements of scale and position angle with and without ADC distortion correction.}
\label{fig:astrometry_hists}
\end{figure}

As a result (see Fig.~\ref{fig:astrometry}), the points fall more closely together both on position angle and scale plots. Without the ADC distortion correction, the standard deviation of the position angle measurements and scale were $0.32^{\circ}$ and $0.12$~mas/px respectively. With the ADC distortion correction, the standard deviation of the position angle measurements and scale are $0.14^{\circ}$ and $0.05$~mas/px, respectively (see Fig.~\ref{fig:astrometry_hists}).

\section{Relative photometry}
\label{sec:photo}

\begin{figure}[h]
\begin{center}
\includegraphics[width=1.0\linewidth]{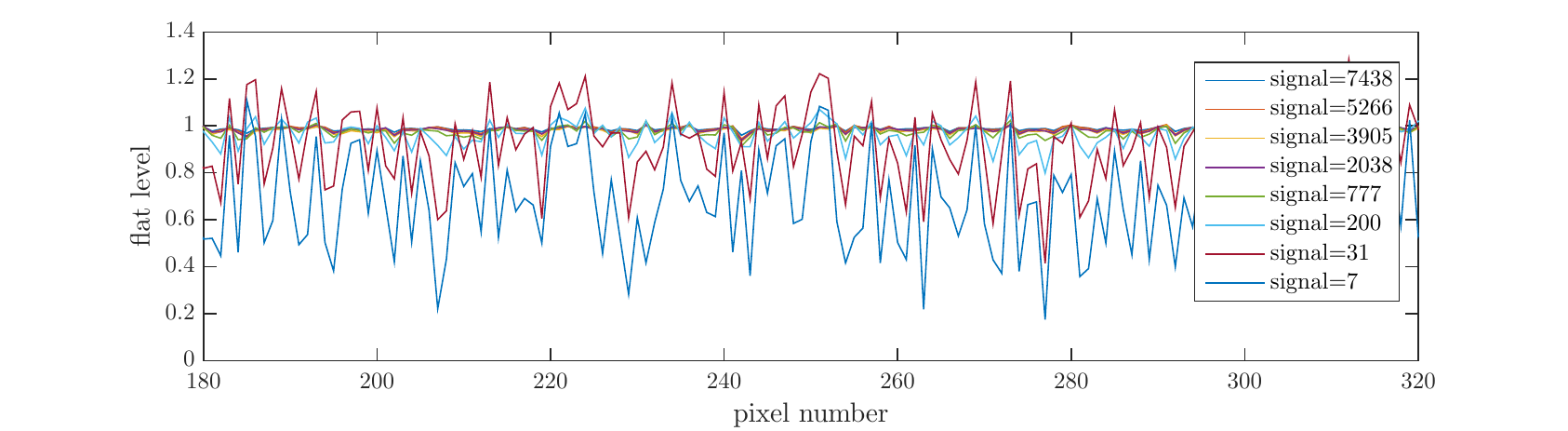}
\caption{Section of flat for different fluxes, as indicated in the legend, in ADU. \label{sec:flatsection}}

\label{fig:flatsection}
\end{center}
\end{figure}

As we discussed before in \citep{Strakhov2023}, the detector demonstrates quite strong non-linearity, especially at low fluxes. Further analysis conducted in the context of flux ratio determination for binary stars showed that exact dependence of signal on flux differs from pixel to pixel. We obtained flat fields during the 1 hour long time interval while the telescope was pointed at the sky during evening twilight. 155000 frames were accumulated in total. Flux level changed from 8000 ADU to 1.5 ADU, which corresponds to 800 and 0.15 electrons, respectively. We then averaged flats in 3000 frames long sections. Fig.~\ref{fig:flatsection} contains a section of normalized flat for different flux levels. There are obvious pixel-to-pixel changes that systematically increase while flux decreases. 

\begin{figure}[h]
\begin{center}
\includegraphics[width=1.0\linewidth]{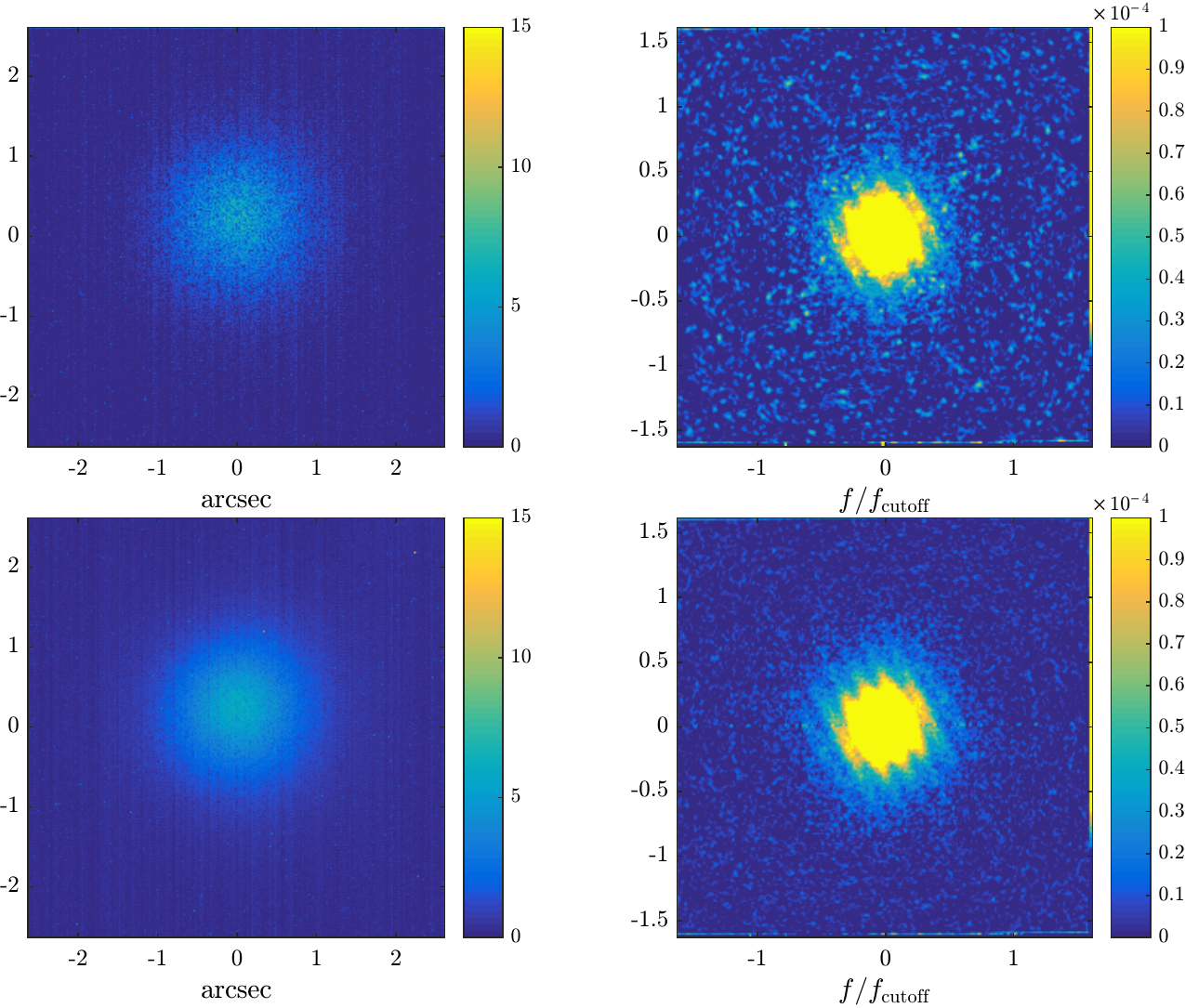}
\caption{Examples of average frame (left column) and average power spectrum (right column). Top row: before flat correction. Bottom row: after flat correction. Flat correction is described in section~\ref{sec:photo}. Average frames are given in ADU, in order to get values in photons one should divide by 10. Power spectrum is normalized by its maximum. Observation of TYC 4281-00675-1 secured on August, 5th 2023. 
\label{fig:img_pspec_demo}}
\end{center}
\end{figure}

Thus, the image of an object is multiplied by some function, which in Fourier space is equivalent to convolution with a filter. This convolution leads to artifacts in the power spectrum. In Fig.~\ref{fig:img_pspec_demo} we demonstrate an average frame and average power spectrum of a faint object before correction for flat. In order to correct for this effect we make interpolation of sky flats averaged to the flux value of a particular pixel. Each frame is divided by pixel--wise interpolated flat. In Fig.~\ref{fig:img_pspec_demo} one can see that after such correction both average image and average power spectra look much smoother.

In order to validate this procedure we conducted a series of observations of 4 binary stars. For this task we have chosen stars representative in the context of survey of TESS targets, having different brightness, separation and flux ratio, see Table~\ref{table:photo_obs}. Each star was observed in different seeing conditions on the following days of 2023 (typical seeing is indicated in parenthesis): August 5th ($\beta\approx1.1^{\prime\prime}$), August 7th ($\beta\approx1.3^{\prime\prime}$), August 27th ($\beta\approx0.55^{\prime\prime}$), August 30th ($\beta\approx0.50^{\prime\prime}$), September 2th ($\beta\approx0.7^{\prime\prime}$), September 5th ($\beta\approx1.1^{\prime\prime}$), September 15th ($\beta\approx1.2^{\prime\prime}$) Also before and after observation of a binary we observed a nearby single star in exactly the same mode. These observations were used in the test of possibility of referencing speckle transfer function (STF), as discussed below. All observations were conducted in $I$ band.

\begin{table}[b]
\begin{footnotesize}
\caption{Binary stars observations for the test of relative flux reproducibility. Meaning of columns: 1. Object. 2. Gaia magnitude. 3--5. Separation $\rho$, mas, Position angle $\theta, ^{\circ}$, flux ratio as measured on August 27th, 2023. 6. Internal error of flux ratio, determined using bootstrap. 7--9. Reproducibility of flux ratio between series obtained on different nights, standard deviation, without flat correction, with flat correction, with flat correction and taking into account speckle transfer function measured using a reference star. 
\label{table:photo_obs}}
\begin{center}
\begin{tabular}[t]{|l|c|c|c|c|c|c|c|c|}
\hline
Object      & G & $\rho$, mas  & $\theta, ^{\circ}$ & f & $\sigma_\mathrm{int}$ & $\sigma_\mathrm{noflat}$ & $\sigma_\mathrm{flat}$ & $\sigma_\mathrm{ref}$\\
\hline
Gaia DR3 2218068072558681088 & 12.6 & $220\pm2$ & $54.4\pm0.4$  & $0.088$  & $0.0057$ & $0.0090$ & $0.0044$ & $0.0073$ \\ 
TYC 3985-1894-1              & 10.4 & $550\pm5$ & $329.2\pm0.4$ & $0.0104$ & $0.0010$ & $0.0046$ & $0.0021$ & $0.0023$ \\ 
TYC 4281-00675-1               & 11.4 & $386\pm3$ & $129.3\pm0.4$ & $0.134$  & $0.0020$ & $0.0026$ & $0.0052$ & $0.0031$ \\ 
Gaia DR3 420954215752408192  & 12.6 & $277\pm3$ & $98.2\pm0.4$  & $0.132$  & $0.0059$ & $0.0286$ & $0.0313$ & $0.0076$ \\ 
\hline

\end{tabular}
\end{center}
\end{footnotesize}
\end{table}

For each observation the parameters of the binary were determined first using standard procedure described in \citep{Strakhov2023}. No flat field correction was applied to frames. Average power spectrum was normalized by its azimuthal average and approximated by the model of binary source. For estimation of internal errors of parameters we employed bootstrapping. The fitting was done for power spectra averaged over 30 subsamples taken randomly from initial series. This value was averaged over observations for one target and presented in Table~\ref{table:photo_obs} as $\sigma_\mathrm{int}$. 

The spread in flux ratio estimations between series $\sigma_\mathrm{noflat}$ is given in 7th column of Table~\ref{table:photo_obs}. One can see that it is significantly larger than internal error $\sigma_\mathrm{int}$. This is how the problem of poor reproducibility of flux ratio measurement manifests itself. The same processing was repeated with flat correction as described above in this section. The 8th column of the same table contains corresponding spread of flux ratio. For stars Gaia DR3 2218068072558681088 and TYC 3985-1894-1 significant improvement is achieved. However, for the remaining two objects the spread is large even with flat correction.

We attribute this abnormally large spread of flux ratio to effect of speckle transfer function (STF). STF is the square modulus of optical transfer function and is influenced by telescope jitter, residual atmospheric dispersion, optical aberrations of the telescope and possibly other factors. The compensation for the STF by normalization of power spectrum by its azimuthal average works improperly when separation of binary is close to diffraction limit and only 1--2 fringes are present in power spectrum.
The problem becomes more severe when the star is faint and the region of Fourier space with decent SNR shrinks. An alternative approach to processing of such measurements is usage of reference stars. Power spectrum of the scientific object is divided by power spectrum of a reference star and then approximated by the model of a binary source. We applied this method to binaries considered in this section. A nearby reference star was observed just before and after each ``scientific'' object. Results of the approximation are presented in the last column of Table~\ref{table:photo_obs}. One can see that for the last three objects the dispersion of flux estimates between series are consistent with internal error of measurement. For the first object (Gaia DR3 2218068072558681088) using reference star is less efficient, probably due to the fact that the reference star was 60 times brighter then the scientific object.

\section{Conclusion}
\label{sec:conc}
In this paper we discuss ways to achieve better precision of astrometric and photometric results of speckle interferometry with upgraded SPP. 

The key way to improve the astrometric results turned out to be the ADC distortion correction. The distortion was measured as the affine geometric transform coefficients using the internal light source. The ADC distortion correction procedure reduced the spread of position angle and separation measurements by more than two times. To estimate this spread, we observed 16 different binaries multiple times to obtain a representative sample. Another important factor taken into account is the telescope collimation error, which we estimate to be equal to $190^{\prime\prime}$. After the correction of the above factors, the resulting standard deviation of the position angle is $0.14^{\circ}$. The relative standard deviation of separation is 0.25\%. 

A closer look at some of observed stars in Fig.~\ref{fig:astrometry}, such as WDSJ02292+5352 or WDSJ03171+4029, hints that there can be some room for further improvement: it is seen that different measurements of the same binary tend to be very close to each other, both on the angle and scale plots. In other words, the spread of measurements of individual binary is much less than the spread of all measurements. This fact can be an evidence of an existence of another source of error. To infer it, a further analysis is required and more observations ought to be obtained to expand the sample.

We tried two approaches to improve the reproducibility of the relative photometry. The first one is to take into account the flux--dependent pixel--to--pixel sensitivity difference, i.e. flux--dependent flat, which is specific for the CMOS detector and not relevant for the EMCCD. This approach resulted in substantial noise reduction in power spectrum hence a slight improvement in relative photometry reproducibility as this noise is partially responsible for errors in flux ratio estimations. The other approach is to compensate the scientific object's visibility function estimation for speckle transfer function measured using nearby single star. If conducted near--simultaneously and close in the sky to the scientific object, this gives us almost an ideal estimation of STF, which automatically includes such factors as telescope jitter, telescope aberrations and residual atmospheric dispersion. The spread in flux ratio between epochs becomes consistent with the internal error of a single measurement.

\acknowledgments 

The authors acknowledge the support of M.V. Lomonosov Moscow State University Program of Development.

\bibliography{references} 

\begin{thebibliography}{10}

\bibitem{Tokovinin2018}
{Tokovinin}, A., ``{Ten Years of Speckle Interferometry at SOAR},'' {\em \pasp}~{\bf 130},  035002 (Mar. 2018).

\bibitem{Ciardi2015}
{Ciardi}, D.~R., {Beichman}, C.~A., {Horch}, E.~P., and {Howell}, S.~B., ``{Understanding the Effects of Stellar Multiplicity on the Derived Planet Radii from Transit Surveys: Implications for Kepler, K2, and TESS},'' {\em \apj}~{\bf 805},  16 (May 2015).

\bibitem{Ziegler2018}
{Ziegler}, C., {Law}, N.~M., {Baranec}, C., {Morton}, T., {Riddle}, R., {De Lee}, N., {Huber}, D., {Mahadevan}, S., and {Pepper}, J., ``{Measuring the Recoverability of Close Binaries in Gaia DR2 with the Robo-AO Kepler Survey},'' {\em \aj}~{\bf 156},  259 (Dec. 2018).

\bibitem{Ziegler2020}
{Ziegler}, C., {Tokovinin}, A., {Brice{\~n}o}, C., {Mang}, J., {Law}, N., and {Mann}, A.~W., ``{SOAR TESS Survey. I. Sculpting of TESS Planetary Systems by Stellar Companions},'' {\em \aj}~{\bf 159},  19 (Jan. 2020).

\bibitem{Scott2021}
{Scott}, N.~J., {Howell}, S.~B., {Gnilka}, C.~L., {Stephens}, A.~W., {Salinas}, R., {Matson}, R.~A., {Furlan}, E., {Horch}, E.~P., {Everett}, M.~E., {Ciardi}, D.~R., {Mills}, D., and {Quigley}, E.~A., ``{Twin High-resolution, High-speed Imagers for the Gemini Telescopes: Instrument description and science verification results},'' {\em Frontiers in Astronomy and Space Sciences}~{\bf 8},  138 (Sept. 2021).

\bibitem{Knudstrup2022}
{Knudstrup}, E., {Serrano}, L.~M., {Gandolfi}, D., {Albrecht}, S.~H., {Cochran}, W.~D., {Endl}, M., {MacQueen}, P., {Tronsgaard}, R., {Bieryla}, A., {Buchhave}, L.~A., {Stassun}, K., {Collins}, K.~A., {Nowak}, G., {Deeg}, H.~J., {Barkaoui}, K., {Safonov}, B.~S., {Strakhov}, I.~A., {Belinski}, A.~A., {Twicken}, J.~D., {Jenkins}, J.~M., {Howard}, A.~W., {Isaacson}, H., {Winn}, J.~N., {Collins}, K.~I., {Conti}, D.~M., {Furesz}, G., {Gan}, T., {Kielkopf}, J.~F., {Massey}, B., {Murgas}, F., {Murphy}, L.~G., {Palle}, E., {Quinn}, S.~N., {Reed}, P.~A., {Ricker}, G.~R., {Seager}, S., {Shiao}, B., {Schwarz}, R.~P., {Srdoc}, G., and {Watanabe}, D., ``{Confirmation and characterisation of three giant planets detected by TESS from the FIES/NOT and Tull/McDonald spectrographs},'' {\em \aap}~{\bf 667},  A22 (Nov. 2022).

\bibitem{Rodriguez2023}
{Rodriguez}, J.~E., {Quinn}, S.~N., {Vanderburg}, A., {Zhou}, G., {Eastman}, J.~D., {Thygesen}, E., {Cale}, B., {Ciardi}, D.~R., {Reed}, P.~A., {Oelkers}, R.~J., {Collins}, K.~A., {Bieryla}, A., {Latham}, D.~W., {Gonzales}, E.~J., {Scott Gaudi}, B., {Hellier}, C., {Jones}, M.~I., {Brahm}, R., {Sokolovsky}, K., {Schulte}, J., {Srdoc}, G., {Kielkopf}, J., {Grau Horta}, F., {Massey}, B., {Evans}, P., {Stephens}, D.~C., {McLeod}, K.~K., {Chazov}, N., {Krushinsky}, V., {Ghachoui}, M., {Safonov}, B.~S., {Dedrick}, C.~M., {Conti}, D., {Laloum}, D., {Giacalone}, S., {Ziegler}, C., {Guerra Serra}, P., {Naves Nogues}, R., {Murgas}, F., {Michaels}, E.~J., {Ricker}, G.~R., {Vanderspek}, R.~K., {Seager}, S., {Winn}, J.~N., {Jenkins}, J.~M., {Addison}, B., {Alfaro}, O., {Anderson}, D.~R., {Aydi}, E., {Beatty}, T.~G., {Bedding}, T.~R., {Belinski}, A.~A., {Benkhaldoun}, Z., {Berlind}, P., {Blake}, C.~H., {Bowen}, M.~J., {Bowler}, B.~P., {Boyle}, A.~W., {Branson}, D., {Brice{\~n}o}, C., {Calkins}, M.~L., {Campbell}, E.,
  {Christiansen}, J.~L., {Chomiuk}, L., {Collins}, K.~I., {Cornachione}, M.~A., {Daassou}, A., {Dressing}, C.~D., {Esquerdo}, G.~A., {Feliz}, D.~L., {Fong}, W., {Fukui}, A., {Gan}, T., {Gill}, H., {Goliguzova}, M.~V., {Hansen}, J., {Henning}, T., {Hintz}, E.~G., {Hobson}, M.~J., {Horner}, J., {Huang}, C.~X., {James}, D.~J., {Jensen}, J.~S., {Johnson}, S.~A., {Jord{\'a}n}, A., {Kane}, S.~R., {Barkaoui}, K., {Kim}, M.-J., {Kim}, K., {Kuhn}, R.~B., {Law}, N., {Lewin}, P., {Liu}, H.-G., {Lund}, M.~B., {Mann}, A.~W., {McCrady}, N., {Mengel}, M.~W., {Mink}, J., {Murphy}, L.~G., {Narita}, N., {Newman}, P., {Okumura}, J., {Osborn}, H.~P., {Paegert}, M., {Palle}, E., {Pepper}, J., {Plavchan}, P., {Popov}, A.~A., {Rabus}, M., {Ranshaw}, J., {Rodriguez}, J.~A., {Roh}, D.-G., {Reefe}, M.~A., {Savel}, A.~B., {Schwarz}, R.~P., {Shporer}, A., {Siverd}, R.~J., {Sliski}, D.~H., {Stassun}, K.~G., {Stevens}, D.~J., {Soubkiou}, A., {Ting}, E.~B., {Tinney}, C.~G., {Vowell}, N., {Walton}, P., {West}, R.~G., {Wilson}, M.~L.,
  {Wittenmyer}, R.~A., {Wittrock}, J.~M., {Wolf}, S., {Wright}, J.~T., {Zhang}, H., and {Zobel}, E., ``{Another shipment of six short-period giant planets from TESS},'' {\em \mnras}~{\bf 521},  2765--2785 (May 2023).

\bibitem{Belinski2022}
{Belinski}, A., {Burlak}, M., {Dodin}, A., {Emelyanov}, N., {Ikonnikova}, N., {Lamzin}, S., {Safonov}, B., and {Tatarnikov}, A., ``{Orbital parameters and activity of ZZ Tau - a low-mass young binary with circumbinary disc},'' {\em \mnras}~{\bf 515},  796--806 (Sept. 2022).

\bibitem{Tokovinin2016}
{Tokovinin}, A., ``{The Triple System Zeta Aquarii},'' {\em \apj}~{\bf 831},  151 (Nov. 2016).

\bibitem{El-Badry2019}
{El-Badry}, K., {Rix}, H.-W., {Tian}, H., {Duch{\^e}ne}, G., and {Moe}, M., ``{Discovery of an equal-mass `twin' binary population reaching 1000 + au separations},'' {\em \mnras}~{\bf 489},  5822--5857 (Nov. 2019).

\bibitem{Malofeeva2023}
{Malofeeva}, A.~A., {Mikhnevich}, V.~O., {Carraro}, G., and {Seleznev}, A.~F., ``{Unresolved Binaries and Multiples in the Intermediate Mass Range in Open Clusters: Pleiades, Alpha Per, Praesepe, and NGC 1039},'' {\em \aj}~{\bf 165},  45 (Feb. 2023).

\bibitem{Strakhov2023}
{Strakhov}, I.~A., {Safonov}, B.~S., and {Cheryasov}, D.~V., ``{Speckle Interferometry with CMOS Detector},'' {\em Astrophysical Bulletin}~{\bf 78},  234--258 (June 2023).

\bibitem{Schlawin2020}
Schlawin, E., Leisenring, J., Misselt, K., Greene, T.~P., McElwain, M.~W., Beatty, T., and Rieke, M., ``{JWST Noise Floor. I. Random Error Sources in JWST NIRCam Time Series},'' {\em The Astronomical Journal}~{\bf 160},  231 (oct 2020).

\bibitem{Primot1990}
Primot, J., Rousset, G., and Fontanella, J., ``Deconvolution from wave-front sensing - a new technique for compensating turbulence-degraded images,'' {\em Journal of The Optical Society of America A-optics Image Science and Vision - J OPT SOC AM A-OPT IMAGE SCI}~{\bf 7},  1598--1608 (09 1990).

\end{thebibliography}
\bibliographystyle{spiebib} 

\end{document}